\documentclass[aps,onecolumn,floatfix]{revtex4}
\usepackage[T1]{fontenc}
\usepackage{graphicx}
\usepackage{amstext}
\usepackage{amsxtra}

\usepackage{amsmath}
\usepackage{amsfonts}

\usepackage{mathtools}

\begin{document}

\title{Spectrally pure single photons at telecommunications wavelengths using commercial birefringent optical fiber}

\author{Jasleen Lugani$^{1}$, Robert J. A. Francis-Jones$^{1}$, Joelle Boutari$^{1}$ and Ian A Walmsley$^1$}

\address{$^1$Clarendon Laboratory, University of Oxford, Parks Road, Oxford, OX1 3PU, UK}

\email{jasleen.lugani@gmail.com} 



\begin{abstract}
We report a bright and tunable source of spectrally pure heralded single photons in the telecom O-Band, based on cross-polarized four wave mixing in a commercial birefringent optical fiber. The source can achieve a purity of 85\%, heralding efficiency of 30\% and high coincidences to accidentals ratio of 108. Furthermore, the possibility of building multiple identical sources is explored. Through the measurements of joint spectral intensities, we find that the fiber is homogeneous over atleast 45 centimeters and can potentially realize 4 identical sources. This paves the way for a cost-effective fiber-optic approach to implement multi-photon quantum optics experiments.
\end{abstract}

\maketitle

\section{Introduction}
Photonics has always been at the forefront of testing fundamental theories of quantum mechanics and optical implementations of various quantum information processing (QIP) protocols~\cite{Kwiat1993,Freedman72,Mandel88,Mandel99,KLM,QC,communication, simulation, metrology}. Building upon a well established technology, it provides avenues to not only generate a wide variety of quantum mechanical states as a resource but also to manipulate them for quantum optical tasks. 
While entanglement is a crucial resource in quantum mechanics~\cite{Nielsen_chuang,10photon_ent}, certain applications require input states to be completely devoid of any correlations~\cite{KLM}. A single photon state is one such example. Ever since the seminal paper on linear optical quantum computing by Knill, Laflamme and Milburn (KLM) in 2001 ~\cite{KLM}, there has been a strong interest in producing single photons with high purity. Furthermore, single photons are also integral to various experiments involving multi-photon interference such as for quantum simulations using photons~\cite{Spring798,bs_qd}. In such tasks, the scale and the complexity of the experiment rely on the availability of multiple indistinguishable single photons with high purity, delivered on demand. 

 One of the most popular ways to generate single photons is through non-linear optical processes such as parametric down conversion (PDC)~\cite{Kwiat95,PRL08_Ian,AE,SPDC_1970,Harder:13} and four wave mixing (FWM)~\cite{Kumar02,Rarity:05,Cohen09,Garay-Palmett:07,Patel_2014,Spring:17,Clark_2011,Francis-Jones:16,Cordier:19}. Although probabilistic in nature, the non-linear optical processes continue to be in widespread use for generating entangled as well as heralded single photons, owing to the simplicity of the experimental setups and tunability in the properties of the photons that can be produced. Furthermore, photon pair generation using FWM in optical fibers allows one to use guided wave dispersion to tailor phase matching conditions such that the resultant bi-photon state is in a factorable state~\cite{PRL08_Ian,Cohen09,Garay-Palmett:07}. Any correlations if present will reduce the purity of the single photon as the detection of the heralding photon will project the conjugate photon in a mixed state. While in recent years, quantum dots (QDs) have emerged as a strong technology to produce on demand single photons~\cite{QD_PS1,QD_PS2}, they need to be operated at cryogenic temperatures. In addition, producing identical single photons from two different QDs is still very challenging.
This issue has been addressed in non-linear processes where generating identical single photons from an array of different waveguides on the same chip has been reported using FWM in doped silica ~\cite{Spring:17}. However, these on-chip waveguides require extensive customization and specialization in fabrication which is not commercial at this stage. The main challenge here is to fabricate identical waveguides with the exact same dimensions and homogeneity as even small differences can lead to distinguishability between the photons generated from separate waveguides. This poses a serious limitation in building multi-photon quantum experiments at large scale. 

We address this issue by exploring FWM in a low cost and commercially available standard birefringent optical fiber (PM980-XP) from Thorlabs to produce single photons in the telecommunications O-Band (1260-1360nm). In comparison to bulk sources and waveguides, photons generated using FWM in optical fibers offer advantages as they are inherently better mode matched to integrated chips and fiber optic platforms that are extensively used in quantum optics experiments. This allows for efficient coupling through these platforms and significantly reducing optical losses through the quantum circuit. With this motivation, in the last few years, there has been a widespread activity in photon pair generation using FWM in different kinds of optical fibers such as photonic crystal fibers~\cite{Fan:05,Cohen09,Goldschmidt08,McMillan:09,Francis-Jones:16}, single mode fibers~\cite{Hall:09,Patel_2014}, dispersion shifted fibers~\cite{Takesue:05,Dyer:09} and also birefringent optical fibers~\cite{Smith:09}. The majority of these approaches rely on tailoring the fiber dispersion and zero dispersion wavelength to achieve phase matching in the desired wavelength region. This is done through customizing the fiber, either by adding some impurities~\cite{Sharping:01} or by changing its transverse spatial structure\cite{Russell:06}. Nevertheless, even through customization, it is still extremely difficult to achieve two fibers with the exact same specifications, which restricts building identical photon sources using them~\cite{Francis-Jones:16}. In contrast to these approaches, we exploit birefringent phase matched FWM in a commercial fiber to produce heralded single photons with high brightness and high purity at telecom wavelengths. Being a standard fiber with a relatively simple structure and a well established fabrication method, it is uniform from draw to draw, and within a single draw, compared to bespoke photonic crystal fibres~\cite{Francis-Jones:16}, allowing one to build multiple sources of identical single photons. It is thus a useful and cost-effective (\$20 per meter) alternative to bulk crystals, bespoke waveguides and fibers in building tunable identical single photon sources. 

While photon pair generation has been earlier explored in birefringent optical fiber~\cite{Smith:09}, the source was not designed for telecom wavelengths and the reported bi-photon state was marked with spectral correlations, thereby limiting the purity of the achievable single photons. In addition, the photons in the idler arm (at 830 nm) were affected by the Raman noise. In this paper, we demonstrate experimentally the potential of a standard birefringent optical fiber as a tunable source of heralded single photons at telecom wavelengths with high brightness, high purity, high heralding efficiency and very low noise. The heralding signal photon at 810 nm opens the possibility for using relatively inexpensive, robust, simple and high efficiency single photon avalanche photodiodes (SPADs) for heralding. Due to the cross-polarized birefringence phase matching, the signal and idler wavelengths are well separated from the pump wavelength, thus greatly reducing the noise due to spontaneous Raman scattering in the idler mode. We substantiate our results by comparing the measured phasematching in the fiber to a theoretical model of the spectral properties of the cross polarized four wave mixing (XFWM) process. The source is further characterized through the measurements of second order coherence and joint spectral intensity using stimulated emission tomography (SET)~\cite{Kwiat2018}.  We also characterize the inhomogenities in the batch of the fiber using the SET technique and explore the possibility of building multiple identical single photon sources using it. Our measurements show that the fiber is relatively homogeneous over a length of 45 centimeter (cm), which can be potentially used to produce four identical single photon sources, with applications in multi-photon quantum optics experiments. 

\section{Cross-polarized four wave mixing (XFWM) enabled by fiber birefringence}
\label{model}
Four wave mixing occurs as a result of the third order non-linear optical response of a medium, in which two photons from a strong pump beam are converted to a signal and an idler photon. The frequencies of the signal and idler photons that can be generated through this process are governed not only from energy conservation but also by the phase matching. In this context, the intrinsic properties of the medium, namely the dispersion as well as the birefringence play an important role to engineer the output state of the signal-idler photon pair~\cite{Garay-Palmett:07}. We consider the XFWM process in which the two photons from a strong pump, polarized along the slow axis of the fiber, give rise to a signal and an idler photon polarized along the fast axis of the fiber. The corresponding energy and phase matching conditions are given by 
\begin{equation}
2\omega_{p}=\omega_s+\omega_i,
\label{ec}
\end{equation}
\begin{equation}
\Delta \beta = 2\beta_{p} - \beta_{s} - \beta_{i} + \frac{2}{3}\gamma P,
\label{pc}
\end{equation}
where $\omega_{p(s,i)}$ and $\beta_{p(s,i)}$ is the angular frequency and propagation constant of the pump (signal, idler) respectively, $\Delta \beta$ is the phase mismatch, $\gamma$ is the third order non-linear parameter of the fiber and $P$ is the peak power of the pump pulses. The last term in Eq. (\ref{pc}) arises from the self (SPM) and cross phase modulation (XPM) of the pump. As the case we consider here is phasematched far detuned from the pump in the normal dispersion regime, and the peak power and nonlinearity of the fiber is relatively weak, we assume that the contribution to the phasemismatch due to SPM and XPM is neglible.  The contribution of the dispersion, both material and waveguiding, and the birefringence of the fiber to the phase matching condition can be explicitly seen by writing Eq. (\ref{pc}) as 
\begin{equation}
    \Delta \beta = 2 \frac{\omega_p}{c}n(\omega_p)-\frac{\omega_s}{c}n(\omega_s)- \frac{\omega_i}{c}n(\omega_i)+ 2 \Delta n \frac{\omega_p}{c} + \frac{2}{3}\gamma P.
    \label{delk}
\end{equation}
Here, $n(\omega_{p(s,i)})$, is the effective refractive index of the fiber at pump (signal,idler) frequency. $\Delta n$ corresponds to the fiber birefringence and is the difference between the effective refractive indices along the two principal axis (fast and slow axis). It must be noted that only those frequencies for signal and idler which simultaneously satisfy Eq. (\ref{ec}) and Eq. (\ref{delk}) are possible. The corresponding bi-photon signal-idler state at the output of the fiber is given by
\begin{equation}
|\psi\rangle=\int \int d\omega_s d \omega_i f(\omega_s,  \omega_i)\hat{a}^{\dagger}_{s}(\omega_s)\hat{a}^{\dagger}_{i}(\omega_i)|0,0\rangle,
\end{equation}
where $\hat{a}^{\dagger}_{s(i)}$ is the creation operator for the signal (idler) mode and 
\begin{equation}
    f(\omega_s, \omega_i)\approx \alpha(\omega_s+\omega_i).\phi(\omega_s, \omega_i),
\end{equation}
is the joint spectral amplitude (JSA) which is a product of pump envelope function $\alpha(\omega_s+\omega_i)$ and phase matching function $\phi(\omega_s, \omega_i)$. In photon pair generation, the joint spectral intensity (JSI), $|f(\omega_s, \omega_i)|^{2}$, is a crucial measurement to understand the state of the generated photon pairs. The shape of the JSI characterizes spectral correlations between the signal and idler photons and is governed by the pump bandwidth $\sigma_p$, the length of the fiber $L$ and the angle $\theta_{si}$ which the phase matching function $\phi(\omega_s, \omega_i)$ makes with the signal frequency axis. The phase matching angle $\theta_{si}$ is dependent upon the group velocities of signal, idler and pump, and thus higher-order dispersion properties of the fiber\cite{Garay-Palmett:07}. For the heralded single photon source, which is the main motivation behind the present work, the respective JSA should be factorable or separable and could be written as a product of signal and idler amplitudes, i.e. $f(\omega_s, \omega_i)=S(\omega_s) I(\omega_i)$. The degree of non-separability determines the degree of spectral correlations present between signal and idler and can be quantified via the Schmidt decomposition of the JSA, where the resulting Schmidt number $K$ determines the number of modes excited by signal and idler photons\cite{Zelinki2018}. $K=1$ corresponds to a factorable state and $K>1$ to a correlated state. The purity of the photons is then defined by $\mathcal{P}=1/K$. Thus, in order to generate photons with high purity, the source should emit photons into a single spectral mode and thus have a factorable JSA. This can be achieved by a careful choice of the pump spectral bandwidth and by engineering the phase matching angle $\theta_{si}$ to be in the range of $[0^\circ$ - $90^\circ]$. 
 
 In order to investigate the JSA of the chosen commercial birefringent fiber, we consider ultrafast pump pulses with Gaussian profile and a bandwidth of 2$\sigma_p$, the pump envelope function can be approximated as
\begin{equation}
\alpha(\omega_s+\omega_i)=\exp\Big[-\frac{(\omega_s+\omega_i-2\omega_{p})^2}{4\sigma_p^2}\Big],
\end{equation}
\begin{figure}[htbp] 
\centering\includegraphics[width = 0.9\textwidth]{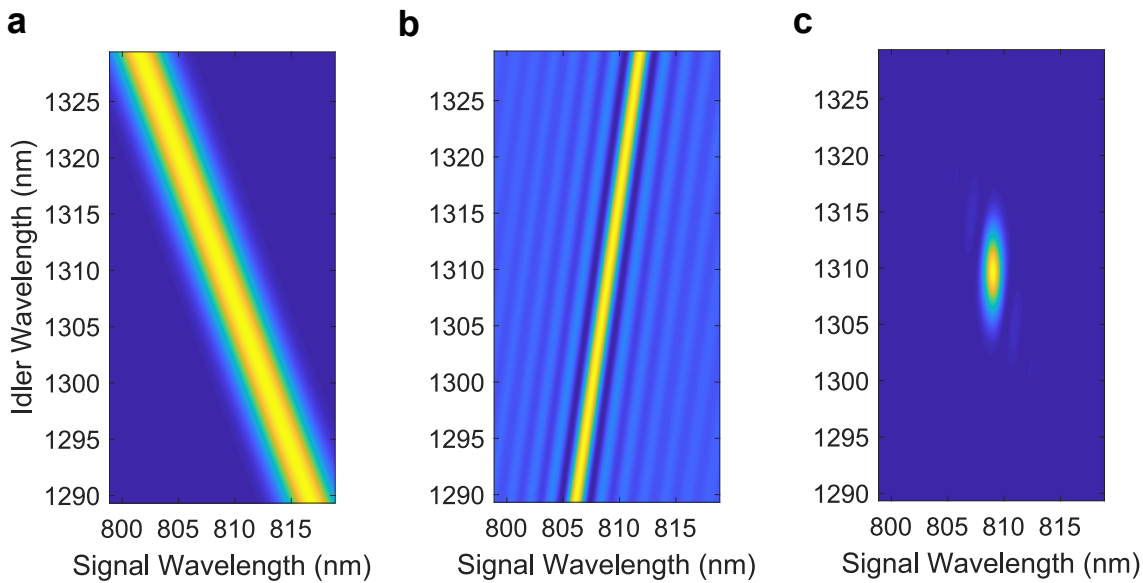}
\caption{Simulations for the FWM process using the model (a) Pump envelope function $\alpha (\omega_s + \omega_i)$ with central wavelength of 1000 nm and bandwidth of 2.5 nm, (b) real part of the phase matching function $\phi(\omega_s,\omega_i$) (c) JSI }
\label{sim_jsi}
\end{figure}
and is plotted as in Fig.\ref{sim_jsi}a. In the present work, we consider the pump with a central wavelength of 1000\,nm and aligned along the slow axis of the fiber to generate a signal photon at 810 nm and an idler photon at telecom (1310\,nm) wavelength, both aligned along the fast axis of the fiber. As can be seen in Fig.\ref{sim_jsi}a, the pump envelope function displays spectral anticorrelation between signal and idler. The phase matching function is given by

\begin{equation}
\phi(\omega_s,\omega_i)= \text{sinc}\big(\frac{\Delta \beta L}{2}\big)\exp\Big(\frac{i\Delta \beta L}{2}\Big),
\end{equation}
where $L$ (= 9 cm here) is the length of the fiber, and $\Delta \beta$ is the phase mismatch given by Eq. (\ref{delk}). $\phi(\omega_s,\omega_i)$ thus depends upon the dispersion and birefrigence of the medium and determines the signal-idler frequencies possible due to phase mismatch over a length $L$ of the fiber. It must be noted that while a standard fiber has a circular core and is isotropic, a birefringent fiber is anisotropic due to the presence of stress rods. This gives rise to difference in refractive indices ($\Delta n$) along its two principal axis. In the present case and in Eq. (\ref{delk}), we model the dispersion of the fiber using a fiber optical toolbox~\cite{OFT} by considering it as a standard circularly symmetric fiber with the given numerical aperture. We then include the effect of fiber birefringence using the beat length value given in the specification sheet of the fiber. In this way, we are able to capture the contribution to the dispersion from the waveguiding structure as well as the material. This is particularly important as the idler field is generated near the zero-dispersion wavelength of the fiber. Following the method described, we evaluate the phase mismatch and thus $\phi(\omega_s,\omega_i)$, which is plotted in Fig.\ref{sim_jsi}b. The width of $\phi(\omega_s,\omega_i)$ is inversely proportional to the length ($L$) of the fiber. As mentioned earlier, the angle $\theta_{si}$ that the JSI makes with wavelength axis (signal, say) provides a good indication of how correlated is the output two photon state. A factorable state is possible when $\theta_{si}$ is in the range between 0 and $90^\circ$, in our scheme, $\theta_{si}= 82^\circ$ (see Fig. \ref{sim_jsi}b). The combination of the pump envelope function and the phase matching function gives rise to the JSI as shown in Fig. \ref{sim_jsi}c. We evaluate the purity $\mathcal{P}$ of the state using the singular value decomposition (SVD) of the JSA and find that for the pump bandwidth $\sigma_p$ (= 2 nm here) and $L$ = 9 cm, a purity of $\mathcal{P}$ = 85\% can be achieved for the output state.
\begin{figure}[htbp] 
\centering\includegraphics[width = 0.5\textwidth]{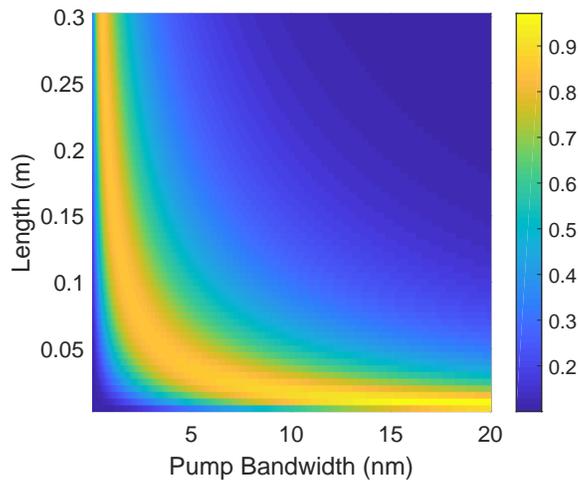}
\caption{Variation of the simulated purity $\mathcal{P}$ of the single photons as a function of fiber length ($L$) and spectral bandwidth $\sigma_p$ of the pump.}
\label{purity_sim}
\end{figure}
It must be noted that $\mathcal{P}$ in this scheme is mainly limited by the experimental choice of $\sigma_p$ and $L$. To demonstrate this, we plot the purity of the output state for a range of pump bandwidths and lengths of the fiber in Fig. \ref{purity_sim}. As evident from the figure, the highest purity is achieved for a length of 1 cm and bandwidth of $\approx$ 17 nm and is around 96\%. However, 1 cm of interaction length in the fiber is too small to give rise to any measurable XFWM. Thus, we have to trade off between the purity of the source and its brightness and chose a length of 9 cm in the fiber and pump bandwidth of 2 nm which give rise to photons with 85\% purity and high brightness as will be discussed in the next section.  

\section{Experiment}
\subsection{Birefringence measurements}
\begin{figure}[htbp]
\centering\includegraphics[width = 0.9 \textwidth]{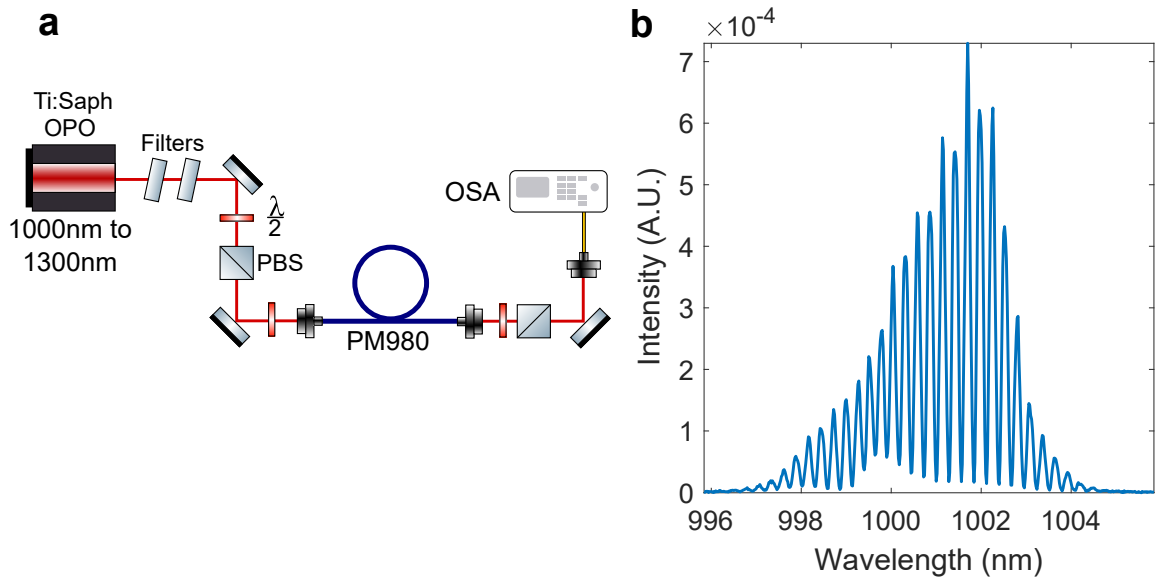}
\caption{(a) Schematic of the experimental setup for measuring fiber birefringence at different wavelengths tuned using an optical parametric oscillator (OPO) and analyzed using optical spectrum analyzer (OSA), (b) measured spectral fringes at the OSA.}
\label{bire}
\end{figure}
 In order to measure the birefringence of this fiber, we sent broadband (around 10\,nm) femtosecond (fs) laser pulses generated via an optical parametric oscillator (OPO) at 80 MHz repetition rate into the fiber at a range of wavelengths including 810\,nm, 1000\,nm and 1310\,nm as shown in Fig. \ref{bire}a. Using a half wave plate (HWP), the light was launched as linearly polarized at 45$^{\circ}$ with respect to the slow axis of the fiber and excited the two orthogonal polarization modes of the fiber. Due to the finite birefringence of the fiber, the difference in the propagation constants of the two orthogonal modes introduces relative phase difference with respect to each other while propagating through it. At the output of the fiber, the light was passed through a polarizer before being detected at an optical spectrum analyzer. In the spectrum, we see fringes as a result of the interference between the two polarization modes in the fiber as shown in Fig.\ref{bire}b. The birefringence of the fiber is related to the fringe spacing according to $\Delta n = \lambda_{0}^{2}/L \delta\lambda$, where $L$ is the length of the fiber. From our measurements, $\Delta n$ was found to be 3.571 $\times$ $10^{-4}$ without any significant change over the wavelengths of operation. This conforms with the value of birefringence calculated using the beat length quoted by Thorlabs's fiber specification sheet. 

\subsection{Photon pair generation using XFWM}
\begin{figure}[htbp]
\centering\includegraphics[width = 0.9\textwidth]{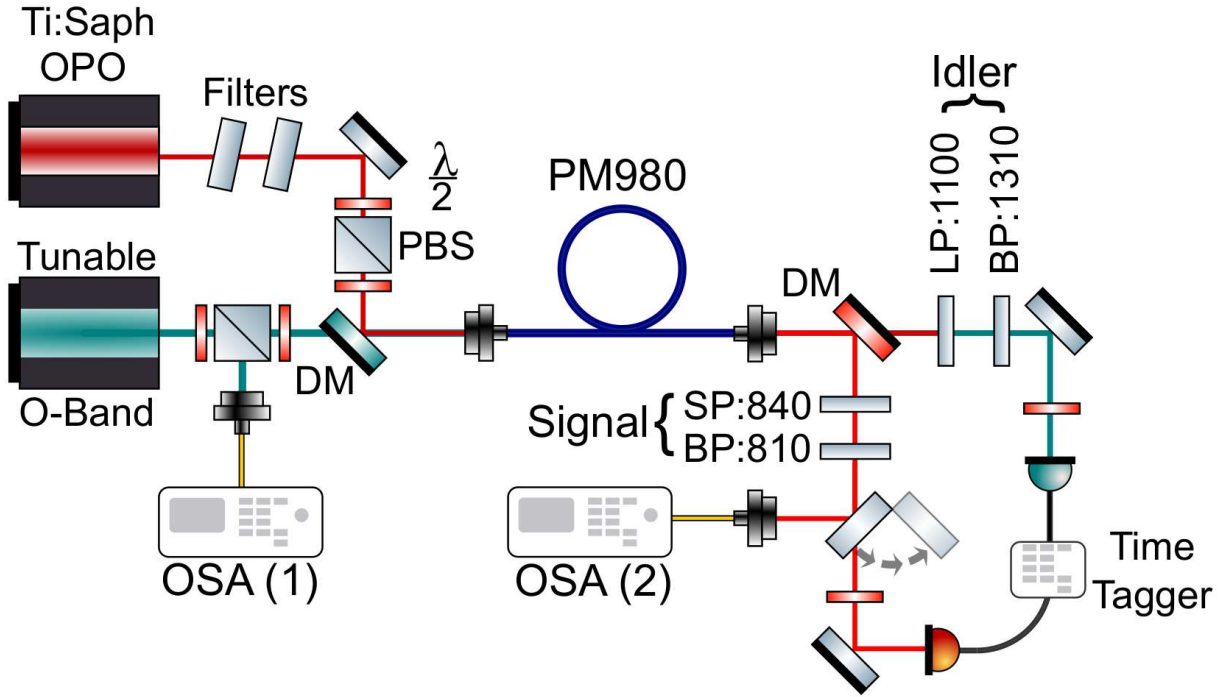}
\caption{Schematic of the experimental setup for realizing cross-polarized FWM to generate heralded single photons at 1310 nm. The pump is derived from a Ti:Sapphire laser driving an OPO resulting in fs pulses at 1000 nm nanometer, which is launched into the FWM fiber (PM980-XP) with the polarization aligned along the slow axis of the fiber. The signal at 810 nm and idler at 1310 nm are separated from each other using a dichroic mirror (DM). The pump is filtered from the generated photons using spectral filters in each arm before collection into single mode fibers. When performing measurements of the JSI, a tunable O-band seed laser was coupled into the fibre together with the pump and is measured on an OSA (OSA(1)). The stimulated FWM signal was then measured for a range of seed wavelengths, on an optical spectrum analyzer (OSA(2)) by diverting the beam to a second collection fiber.} 
\label{fig:scheme}
\end{figure}

The experimental setup used to realize the fiber source is shown in Fig. \ref{fig:scheme}. The pump is obtained from a Ti:Sapphire laser which produces femtosecond pulses at 780 nm and drives an OPO, yielding broadband (30 nm, FWHM) femtosecond laser pulses with central wavelength 1000 nm and 80 MHz repetition rate. The pump bandwidth is reduced to $\approx2.5$\,nm (FWHM) by using a combination of spectral filters - long pass edge filter (LP02-808), angle tuned hard edge filter (LL01-1030) and short pass filter (TSP01-1116). The input pump power can be varied between 0 and 70 mW using a polarizing beam splitter (PBS) and a half waveplate (HWP). The pump is launched into the fiber ($L$ = 9 cm) using an aspheric lens. An additional HWP after the PBS controls the polarization of the pump before it enters the fiber and aligns it along the slow axis of the fiber. The pump in this configuration results in signal and idler photons at 810 nm and 1310 nm, orthogonally polarised to the pump, i.e. along the fast axis of the fiber. The pump, signal, and idler are coupled out of the fibre using another aspheric lens.

 Due to the large difference in wavelength between the signal and idler photons, it is not possible to collimate both signal and idler beams simultaneously. Instead, the heralding signal mode was prioritized in order to maximize the number of potential heralding events. The signal and idler modes were separated using a dichroic mirror (DM) with an edge at 950\,nm. The pump suppression in each of the arm was performed using spectral filters - a bandpass filter (810-DF10) and short pass filter (FF01-842/SP) in the signal arm and a long pass filter (BLP01-1064) and band pass filter (BP:1310) at 1310 in the idler arm. The signal and idler were coupled into single mode optical fibers and detected using superconducting nanowire detectors (SNSPDs) which have a detection efficiency of more than 85\% for signal and 50\% for idler. 

\section{Results}
\subsection{Phasematching contours}
\begin{figure}[htbp]
\centering\includegraphics[width = 0.9\textwidth]{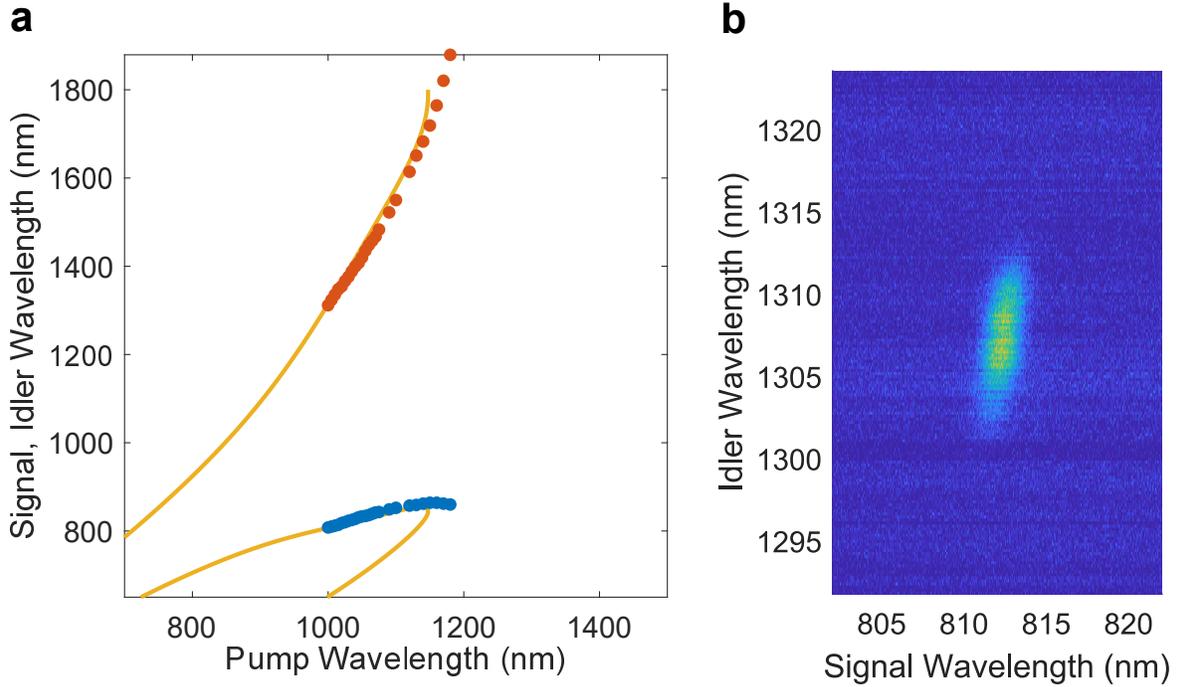}
\caption{(a) Phase matching contours - theoretical model (solid line) and experimental measurements (dotted points) (b) Measured JSI using seeded FWM}
\label{phasecontours}
\end{figure}

 In order to explore the birefringence phase matching in the fiber, we measured the perfectly phase matched signal-idler frequencies for a range of pump wavelengths. To do this, we tuned the pump wavelength from 1000 nm to 1100 nm by scanning the OPO and recorded the spectrum of the signal photons using a single photon level spectrometer (Andor Solis Shamrock 193i). The frequency of the conjugate idler photon was evaluated using energy conservation (Eq. \ref{ec}).  The corresponding results are plotted in Fig.\ref{phasecontours}a as scatter points, the blue and red corresponds to the wavelengths of the recorded photons in the signal arm and idler arm respectively. The solid curve corresponds to the phase matched contours plotted using the theoretical model described in Sec. \ref{model}. As evident in Fig.\ref{phasecontours}a, the experimental measurements are in good agreement with the phase matched contours obtained using the model. 
Since the JSI is a good measure to explore the output bi-photon state, we carry out a seeded measurement which stimulates the FWM process and results in difference frequency generation (DFG) \cite{Sipe:13,Francis-Jones:16}. As both the processes occur via the same phase matching condition, the stimulated signal directly corresponds to the probability amplitude of the spontaneously generated signal-idler photon state. Since the spontaneous process is inherently weak, the seeded FWM offers advantages by generating a much stronger signal which can be easily recorded using a standard OSA. To carry out the JSI measurement, we fix the pump wavelength at 1000 nm and launch it in the fiber along with a tunable continuous wave seed laser which can tune between 1285 nm and 1325 nm. The polarization of the seed laser is aligned orthogonal to the pump laser, i.e. along the fast axis of the fiber. We record the signal generated through DFG by collecting it in the signal arm and sending it to an OSA. The wavelength of the seed laser was scanned and the corresponding signal is recorded for each scan. The wavelength of the seed laser at each set-point was also recorded on a second OSA. This is needed as the seed laser is not well calibrated nor is its setting repeatable. Therefore we must re-calibrate the vertical axis of each JSI recorded by using this extra data. The resulting JSI measurement is shown in Fig. \ref{phasecontours}b. The measured JSI is in good agreement with the one predicted from the model (Fig. \ref{sim_jsi}c). From the recorded JSI, we make an estimate of the JSA (assuming flat spectral phase) by taking the square root. To this approximate JSA, we apply a singular value decomposition to determine the approximate number of modes present and hence determine an upper bound on the purity of $\mathcal{P} = 75$\%.

\subsection{Source characterization}
\begin{figure}[htbp]
\centering\includegraphics[width = 0.95\textwidth]{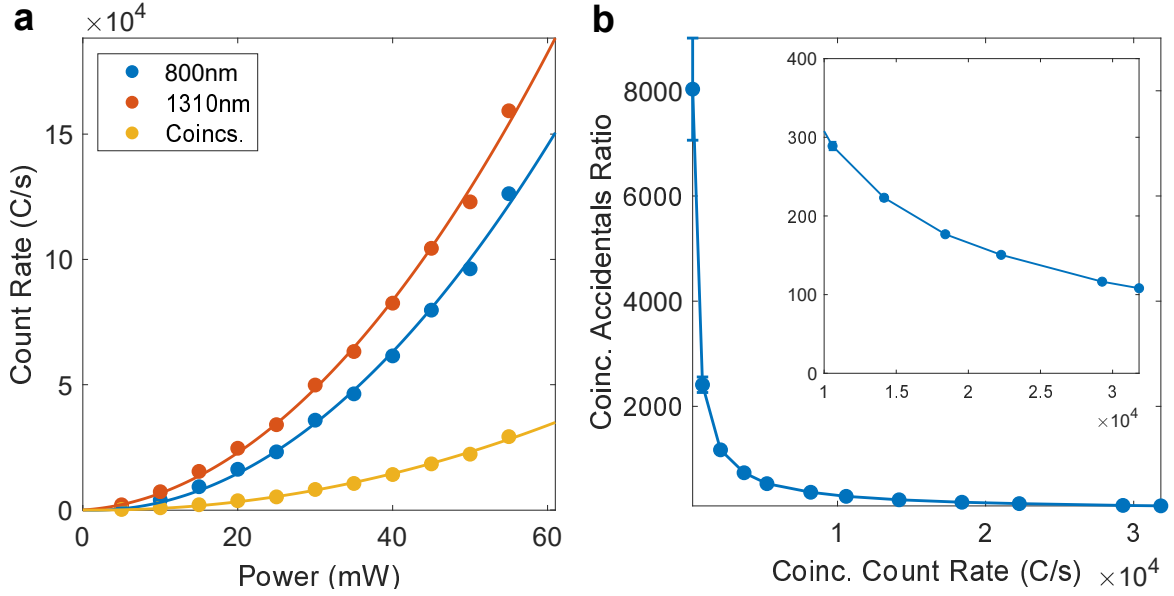}
\caption{(a) Variation of the measured count rates for signal (red) and idler (blue) photons and coincidences (yellow) and the quadratic fits, (c) Variation of Coincidences to Accidentals Ratio with pump power.}
\label{crate}
\end{figure}

The phase matching and JSI measurements above (Fig. \ref{phasecontours}) illustrate that it is possible to generate a factorable bi-photon state with one of the photons at 810 nm which can be used to herald the presence of the other photon at 1310 nm, with high heralding efficiency and purity. For further characterisation of the source, we collect the signal and idler photons in to their respective single mode fibers (Fig. \ref{fig:scheme}). The collection fibers are then coupled to the respective SNSPDs. The detection signals from SNSPDs were fed to a time-tagger with 81ps resolution in which coincidence detection was performed. We varied the input pump power and measured the count rates at the two detectors ($N_{s}$, $N_{i}$) and the rate of coincident detection events ($N_{si}$) between the two. For the parametric XFWM process, the number of signal and idler photon pairs generated should increase quadratically with increasing pump power, i.e. $N_{si} \propto P^2$. This is clearly exhibited in the measurements as shown in Fig. \ref{crate}a. With optimization, we could achieve a maximum count rate of  $N_{si}=30,000\ C/s$ for a 9 cm long fiber, with a heralding efficiency $\eta$ between 20$\%$-30$\%$ for both signal and idler arms. The data presented throughout this manuscript has been background subtracted and the error bars have been calculated assuming Poissonian photon counting statistics. It must be noted that the signal and idler photons are free from any Raman noise as is evidenced from the nearly pure quadratic variation of the count rates with pump power (Fig. \ref{crate}a).  In order to confirm the XFWM process, we rotated the polarization of the pump by $90^\circ$, i.e. to align it along the fast axis of the fiber, which dropped the signal and idler counts to zero, thus eliminating the XFWM process. We also measure the coincidences to accidental ratio (CAR) for the source which is given by
\begin{equation}
    CAR=\frac{N_c}{A_c}=\frac{N_{si} R_p}{N_s N_i},
    \label{car}
\end{equation}

 where $N_{si}$ are the coincidence counts and $R_p$ is the repetition rate of the laser. In Fig. \ref{crate}b, we plot the measured CAR as a function of the achieved coincidence count rate between the signal and idler modes. A CAR > 1 implies the existence of significant photon number correlation between the two arms of the source. In the literature concerning the development of fiber based single-photon sources, it has been suggested that a CAR in excess of 10 demonstrates the potential usefulness of source for use in quantum technologies~\cite{Migdall_book}. In comparison to the earlier reported values \cite{McMillan:09,Francis-Jones:16,Migdall_book,Tanzilli_LaserP_review} for fiber based single photon sources, we have measured much higher CAR, > 108 over the whole range of pump powers. This high value can in part be attributed to the reduction of the spurious singles counts arising from spontaneous Raman scattering in the idler mode, that has been made possible by operating under a  birefringent phase-matching scheme. This value of the CAR could be increased further through the addition of simple optical switch to optically gate the idler photons on the condition of a herald being detected~\cite{Francis_Jones_2017}.

\begin{figure}[htbp]
\centering\includegraphics[width = 0.95\textwidth]{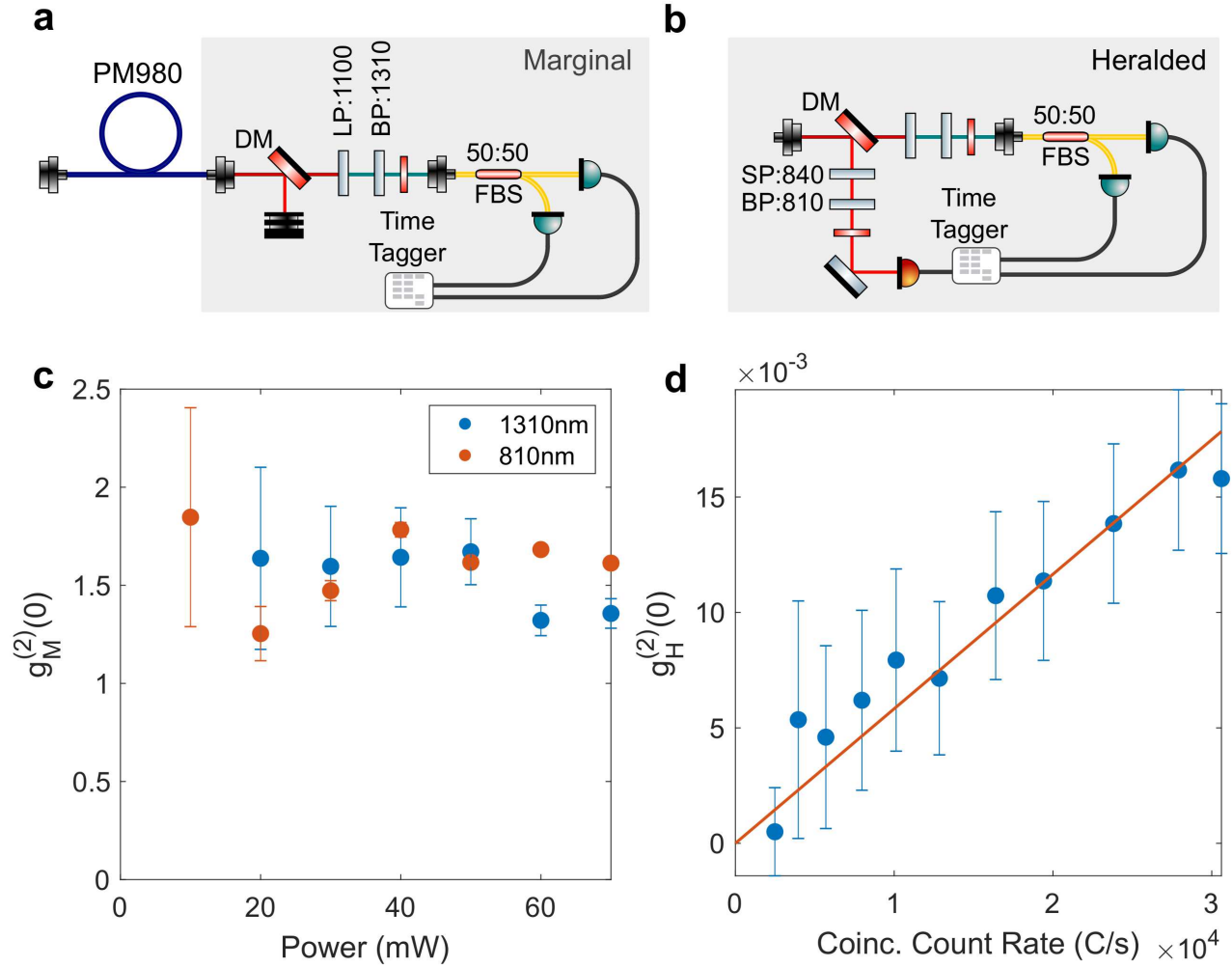}
\caption{(a) Detection scheme for measuring the marginal second order coherence of one arm, in this case the idler, (b) Detection scheme for measuring the heralded second order coherence of the idler arm conditioned on the detection of a heralding photon in the signal arm. (c) Variation of the marginal $g^{(2)}_{m}(0)$ for the signal (red circles) and idler (blue circles) as a function of pump power and (d) $g^{(2)}_h(0)$ for the idler (blue circles) as a function of the coincidence count rate of the source, red line is a linear fit to the data as a guide to the eye.}
\label{g2s}
\end{figure}

Another metric of interest to quantify the purity of heralded single photons is the second order coherence $g^{(2)}(\tau)$, where $\tau$ is the time interval between the detections~\cite{Loudon}. We measure $g^{(2)}(0)$ for both signal (810 nm) and idler (1310 nm) photons separately and call it marginal $g^{(2)}_{m}$. The value of marginal $g^{(2)}_{m}(0)$ is sensitive to the single photon purity as it strongly depends on the number of Schmidt modes in the two photon state ~\cite{Andreas-Eck:11,Kwiat2018} and relates as  $g^{(2)}_{m}=1+\frac{1}{K}$, where $K$ is the number of modes ~\cite{Migdall_book}. In an ideal case, when there is only one Schmidt mode ($K=1$), photons in each of the arm have thermal statistics, giving a marginal $g^{(2)}_{m}(0)$ of 2. As the source departs from being ideal and emits into multiple Schmidt modes, the value of $g^{(2)}_{m}(0)$ reduces from 2 tending towards a value of 1 for a highly multimode state. Thus, measuring $g^{(2)}_{m}(0)$ gives information about the multi-modedness ($K$) and in turn, the single photon purity as $\mathcal{P}=1/K$ . In order to experimentally measure $g^{(2)}_{m}(0)$, we couple photons in the respective arms in a 50:50 fiber beam splitter (FBS), the output ports of which are monitored using SNSPDs (Fig. \ref{g2s}a). The singles counts ($N_{d1}$) and ($N_{d2}$) and coincidence counts ($N_{dc}$) are integrated over 10 seconds. The marginal $g^{(2)}_{m}(0)$ is then evaluated as
\begin{equation}
    g^{(2)}_{m}(0)=\frac{N_{dc}R_{p}T_{int}}{N_{d1}N_{d2}},
\end{equation} 
where $R_p$ is the repetition rate of the laser (80 MHz) and $T_{int}$ is the integration time of 10s. The marginal $g^{2}_{m}(0)$ for both signal and idler photons is plotted as a function of pump power in Fig. \ref{g2s}c. As can be seen the value of marginal $g^{(2)}_{m}(0)$ for both signal and idler remains fairly constant over the range of pump powers used, and has a maximum of $1.71\pm 0.2$ for the idler (1310\,nm) arm and $1.916\pm0.39$ for the signal (810 nm) arm, which gives a purity of 75\% and 91\% respectively. 

We have also measured the second order coherence conditioned on the presence of a heralding photon in the signal arm ($g^{(2)}_{h}(0)$). In this case, we measure the rate of three fold coincidences $N_{s,i1,i2}$, as well as the coincidences between the heralding detector and either of the remaining idler detectors, $N_{h,i_{1}}$ and $N_{h,i_{2}}$. The $g^{(2)}_{h}(0))$ is a useful measure of the contamination of the bi-photon state due to the emission of higher order photon-pair events~\cite{URen:05,Goldschmidt08} and can be written as
\begin{equation}
    g^{(2)}_h(0)=\frac{N_{h,i_1, i_2}N_h}{N_{h,i_{1}}N_{h,i_{2}}}.
\end{equation}
For an ideal single photon source $g^{(2)}_h(0)=0$, indicating that whenever there is a photon detected in the signal arm, there is only one idler photon and the probability of detecting more than one idler photon is zero. We measure $g^{(2)}_h(0)$ for different pump powers and plotted against the signal and idler coincidence count rate as shown in Fig. \ref{g2s}d. The heralded $g^{(2)}_h(0)$ remains $<<1$ for all pump powers used, indicating the heralded single photon state is nonclassical. As the pump power increases, $g^{(2)}_h(0)$ increases linearly with the coincidence count rate as the probability of multi-photon events increases. However, even for the largest operating pump power (70 mW), the measured $g^{(2)}_h(0)=0.017\pm0.004$ remains low, indicating a very low level of multi-photon or any other spurious noise (residual pump, Raman scattering etc.) contamination. 

\section{Multiple identical sources}
\subsection{Testing for fiber inhomogenity}
\label{fiber_inhomogenity}
\begin{figure}[htbp]
\centering\includegraphics[width = 0.9\textwidth]{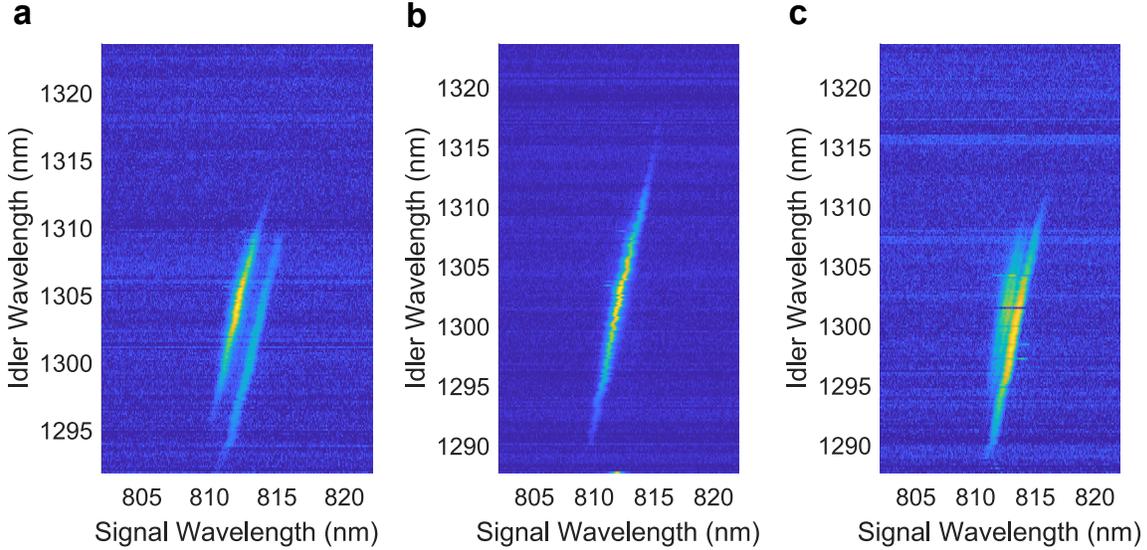}
\caption{JSI measurements for different lengths of the fiber: (a) 1 meter (b) 45 centimeter (c) 20 centimeter}
\label{inhomogenties}
\end{figure}
Motivated by the source performance, we further explored the possibility of building multiple identical sources using the same fiber. Since birefringence is a key aspect in this source, for developing identical sources, it is important that the respective fibers have the same birefringence. For a commercial fiber such as PM980-XP, the manufacturing is much simpler, and well established in comparison to bespoke fibers such as photonic crystal fibers. We therefore expect it to be more homogeneous along its length within a single draw and from batch to batch. The inhomogenities in the fiber such as modulations in the core size (or impurities) or different magnitudes of stress induced by the stress rods in the cladding, if present, will change the birefringence of the fiber in that region resulting in a different effective refractive index and thus alter the phase matching conditions. This will result in spectral distinguishability between photons generated from two different pieces of fiber taken from the same batch and will limit their use in multi-photon interference experiments~\cite{Francis-Jones_PCF_uniformity,Cui:12}. In this work, through JSI measurements, we have experimentally investigated the inhomogenities of a batch of a fiber for building multiple identical photon sources. For this, we took 1 meter of fiber and measure the JSI using the SET technique as shown in Fig. \ref{inhomogenties}a. As evidenced in the figure, the JSI has multiple features, which is due to the fiber having different birefringence values over its length, each giving rise to different phase matching as well as XFWM bandwidths. To understand the length scale of the inhomogenity, we cut the fiber into two pieces, 45 cm each and repeat the JSI measurements for both. The respective JSI are shown in Fig. \ref{inhomogenties}b,c. While the JSI of one half still exhibit multiple features (Fig. \ref{inhomogenties}c), the JSI for the other half (Fig.\ref{inhomogenties}b) exhibit only one feature corresponding to a single phase matching indicating that the fiber is homogeneous throughout this length. This is further confirmed by cutting the fiber further into two 20 cm pieces which again gives rise to uniform JSIs. Thus, the JSI measurements show that the fiber is homogeneous over atleast  a length of 45 cm, indicating the potential use of building multiple sources using the same batch of fiber. 

\subsection{Identical fiber sources}
\begin{figure}[htbp]
\centering\includegraphics[width = 0.65\textwidth]{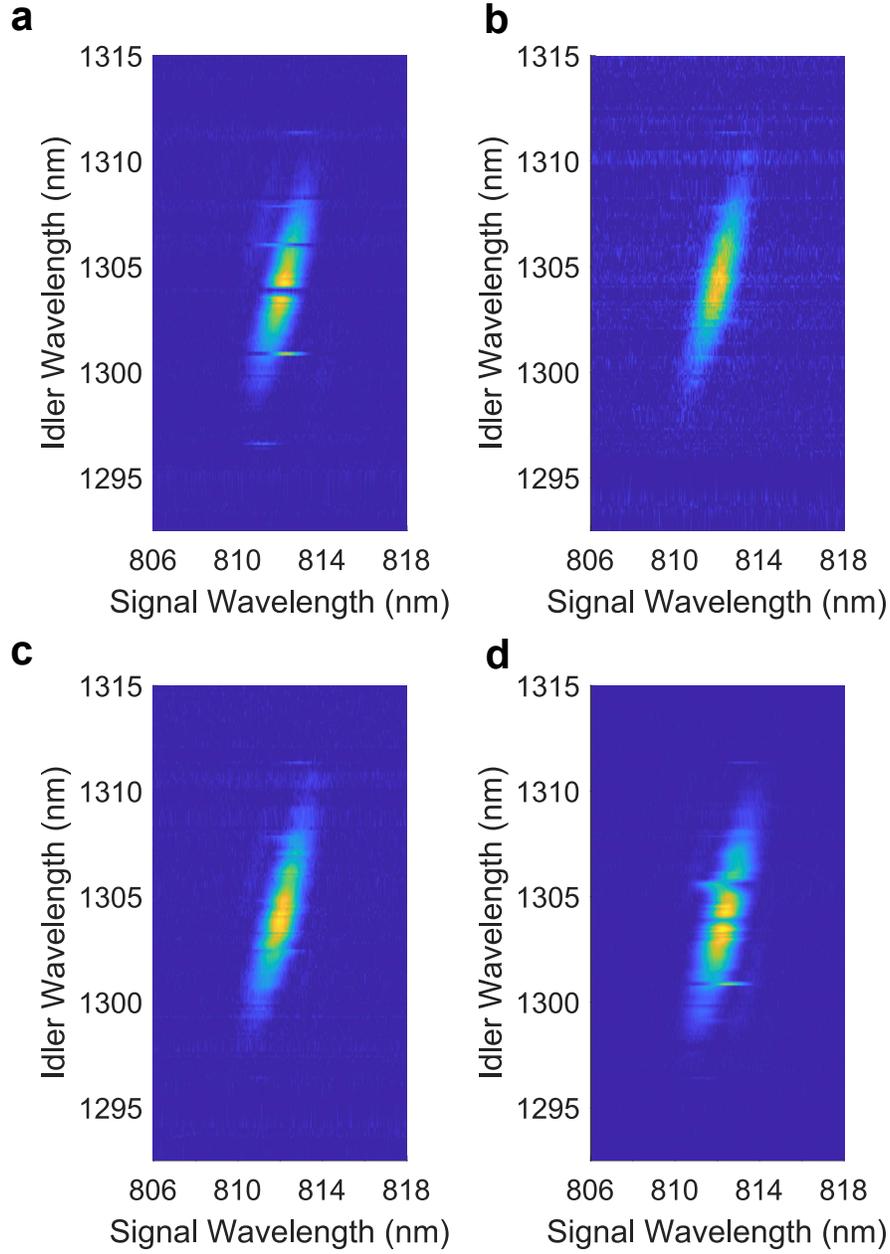}
\caption{Experimentally measured identical JSIs of 4 different fibers (a-d), $L$= 15 cm each. The pump bandwidth has been increased to improve the measureable signal level. As a result these JSIs are no longer fully factorable, however with the correct pump filtering in place a factorable state would be achieved.}
\label{idnfib}
\end{figure}
We began with 60 cm of the fiber PM980-XP from the same batch, and cut it into 4 pieces, each 15 cm in length. We performed the JSI measurements of each piece using the SET technique. The results are shown in Fig. \ref{idnfib}, the respective JSIs look quite identical with an overlap ranging between 85\% to 91\%. In order to improve the signal levels, we increased the bandwidth of the pump to increase the available power from the OPO, as a result one should not expect these to appear factorable. However, the overlap will still remain high and may even improve with the correct pump filtering imposed to form a factorable two-photon state. It must be noted the overlaps could be improved by using a different tunable seed laser for the idler as the used laser has issues with the stability and calibration. As stated above, we have calibrated the vertical axis of each JSI by recording the actual value of the laser at this set-point, however there are still some discrepancies in the JSI as the seed wavelength is swept. This results in differences in the JSIs from one scan to another and attributes to a lower value of the overlap. To determine the overlap, we first calibrate the idler wavelength axis, following this we perform a 2D interpolation of the data so that all four data sets exist on a single mesh of signal and idler wavelengths, from this the overlaps were evaluated by normalising each JSI, and then calculating the approximate JSA through the square root. The overlaps of all possible combinations was then performed by integrating the product of a pair of JSAs over the signal and idler wavelengths.
\subsection{Heralded single photons in telecom C band}
\begin{figure}[htbp]
\centering\includegraphics[width = 0.85\textwidth]{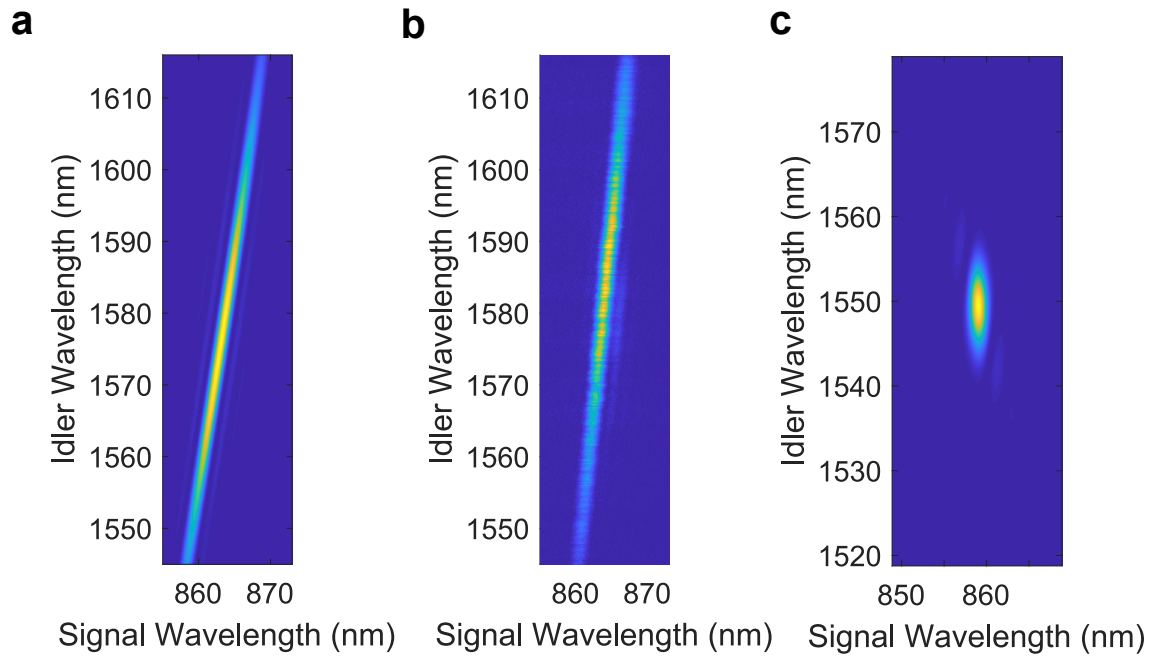}
\caption{(a) Theoretical and (b) Measured JSIs corresponding to pump tuned to 1110 nm and idler photons at 1550 nm.}
\label{jsi1550}
\end{figure}
In the present work, we specifically focused on generating heralded single photons at 1310 nm for applications in time-bin quantum optics experiments using large fiber-optic networks where low dispersion is a key requirement. We have further explored the possibility of using the same scheme and fiber to possibly generate idler photons in the telecommunications C-Band. For this we tuned the pump to 1110 nm and the seed laser from 1540 nm to 1620 nm and measured the seeded JSI as shown in Fig. \ref{jsi1550}a. As with the homogeneity study, we removed all pump filtering, with the exception of a longpass filter to suppress any remaining Ti:Sapph pump light exiting the OPO at 810nm. In doing so, we were able to achieve higher signal levels as a result of the increased pump power, but at the expense of the shape of the JSI. Nevertheless, we see good agreement between our experimental data, and the JSI predicted by our numerical model of the fiber phasematching with an appropriate pump envelope function applied. This does let us clearly observe the angle of the phasematching function relative to the signal axis, as the orientation of the JSI is dominated by the bandwidth of the phasematching in the fiber. High factorability can be restored by appropriately filtering the pump as see in Fig.\,\ref{jsi1550}c.

\section{Conclusion}
In this paper, we have demonstrated a tunable and bright heralded single photon source using  XFWM in a low-cost, commercially available birefringent fiber. We have performed a thorough  characterization of the source including measurements of the join spectral intensity and second order coherence, all of which suggest that the source generates single photons with high purity and low noise evidenced by a CAR$> 100$ and $g^{(2)}_{h}(0) < 0.02$. The simple theoretical model of the fiber and birefringent phase matching simulates the phase matched wavelengths and JSIs with good agreement to the experimental measurements. Whilst we have utilised a Ti:Sapphire pumped OPO as the source of our pump light, the combination of wavelengths involved and the phasematching of the fibre, also make this a good candidate for constructing heralded single photon sources from cheaper and more compact Ytterbium doped fibre lasers, leading to the potential of a fully fibre integrated package of pump and photon source in a small footprint.

Furthermore, we have explored the possibility of building multiple identical sources using the same batch of fiber. In order to build such sources, the respective fibers should be identical in all their aspects - length, core size and birefringence, implying the fiber should be uniform and homogeneous over the usable length. We have used the SET technique to determine the structural inhomogenities of a small sample in a batch of fiber. 
Our measurements show that the commercial fiber is relatively homogeneous over a length of 45 cm, and thus has the potential to build at least 4 single photon sources with the performance metrics specified in this manuscript. While the work presented here focused on generating heralded single photons at 1310 nm for applications in quantum optics experiments involving long fiber-optic networks where low dispersion is a key requirement, we have also investigated the tunability of the source for producing photons at 1550 nm. 
This opens up avenues to use this tunable and cost effective source for a wide range of applications in quantum optics experiments, both in multi-photon interference experiments in long fiber-optic networks or large scale integrated photonics ciruits. 

\section*{Acknowledgments}
The authors thank Steve Kolthammer and Thomas Hiemstra for fruitful discussions.

\section*{Funding}
UK's National Quantum Technologies Programme Networked Quantum Information Technologies (NQIT) hub under Grant No. EP/N509711/1; the European
Commission H2020-FETPROACT-2014 grant QUCHIP; ESCHER (EP/R041865/1); ERC (Advanced Grant MOQUACINO)


\bibliography{sample}

\end{document}